\documentclass[a4paper,11pt]{article}
\usepackage{pos}
\usepackage{color}
\usepackage{comment}
\usepackage{amsmath}
\usepackage{slashed}

\definecolor{todocolor}{RGB}{255,100,100}
\newcommand{\todo}{\colorbox{todocolor}{\textsc{Todo}}}

\DeclareMathOperator{\Tr}{Tr}

\title{Real-Time Dynamics At Large $N$}

\author*{Scott Lawrence}

\affiliation{Department of Physics, University of Colorado,\\
  Boulder, CO 80309, USA}

\emailAdd{scott.lawrence-1@colorado.edu}

\abstract{The large-$N$ limit of $O(N)$-symmetric bosonic field theories, or $U(N)$-symmetric fermionic field theories, is amenable to a saddle point approximation. As a result, there is a family of closely related algorithms for efficient lattice simulations in this limit, even in the presence of fermionic or real-time sign problems. These can be used to study quenches, or other observables for which analytic large-$N$ calculations become impractical.}

\FullConference{%
 The 38th International Symposium on Lattice Field Theory, LATTICE2021
  26th-30th July, 2021
  Zoom/Gather@Massachusetts Institute of Technology
}

\begin{document}
\maketitle

\section{Introduction}

The dynamics of thermal field theories are well described, in the long-time and large-distance limits, by relativistic hydrodynamics~\cite{Romatschke:2017ejr}, which has the equation of state and transport coefficients (chiefly the shear and bulk viscosities) as its low-energy constants. While lattice Monte Carlo methods are supremely effective at calculating the equation of state (at least for fermion-antifermion symmetric systems), they are precluded by a sign problem from directly accessing real-time dynamics. As a result, reliable calculations of the shear viscosity $\eta$ of QCD are still unavailable~\cite{Moore:2020pfu}, although methods based on the reconstruction of the spectral function~\cite{Meyer:2011gj} provide some estimates, compatible with experimental results, that the ratio of shear viscosity to entropy density has a minimum of little more than $\left(\frac \eta s\right)_{\mathrm{KSS}} \sim 0.08$.

In certain limits, the shear viscosity (and other transport coefficients) may be obtained analytically. The most famous such result is the holographic calculation of $\mathcal N=4$ supersymmetric Yang-Mills in the strong coupling and large-$N$ limits~\cite{Policastro:2002se}. Similar (albeit arguably more difficult) calculations have been performed in the weak-coupling~\cite{Jeon:1994if} and large-$N$~\cite{Romatschke:2021imm} limits of scalar field theory.

The procedure used to obtain large-$N$ results from scalar field theory is chiefly analytic, but it can be used just as well as a saddle point expansion on the lattice. In this form it is a close relative of the Lefschetz thimble methods for evading the sign problem~\cite{Cristoforetti:2012su}. This talk will describe several algorithms for working in the large-$N$ limits of bosonic and fermionic field theories, and demonstrate their operation in low dimensions. An important advantage of these methods is that they give ready access to observables that are otherwise not analytically tractable; for instance, the study of quenches.

\section{Bosonic and Fermionic Models}
The first model we will target is the $N$-particle generalization of the quartic anharmonic oscillator. The $i$th particle has position operator $x_i$ and canonically conjugate momentum $p_i$; the full system is described by the Hamiltonian
\begin{equation}\label{eq:H_aho}
H_{AHO} = \sum_i \frac 1 2 \left( m^2 x_i^2 + p_i^2\right) + \lambda \left(\sum_i x_i^2\right)^2\text.
\end{equation}

Time-dependent expectation values in a thermal state, of the form $\langle \mathcal O(T) \mathcal O(0)\rangle$, can be expressed in a path integral formulation by beginning with one of the definitions
\begin{equation}\label{eq:cor}
\langle \mathcal O(T) \mathcal O(0)\rangle
\equiv
\Tr e^{-\beta H} e^{i H T} \mathcal O e^{-i H T} \mathcal O
\equiv
\Tr e^{-\beta H / 2} e^{i H T} e^{-\beta H / 2} \mathcal O e^{-i H T} \mathcal O
\end{equation}
or another equivalent definition,
and expanding the exponentials of the Hamiltonian via the Trotter-Suzuki approximation. This standard procedure (see~\cite{Alexandru:2016gsd} for a detailed discussion) yields the action
\begin{equation}\label{eq:S_aho}
S_{AHO} = \sum_{t,i} \frac{(\phi_i(t+1)-\phi_i(t))^2}{2 a(t)} + \sum_{t} \frac{a(t) + a(t-1)}{2} \left[\frac{m^2}{2} \sum_i \phi_i(t)^2 + \lambda \left(\sum_i \phi_i(t)^2\right)^2\right]
\text.
\end{equation}
Here $a(t)$ defines the so-called ``Schwinger-Keldysh'' contour, representing both imaginary- and real-time evolution. The order in which imaginary- and real-time evolution operators are applied in (\ref{eq:cor}) determines $a(t)$.

\begin{figure}
\hfil
\includegraphics[width=0.47\linewidth]{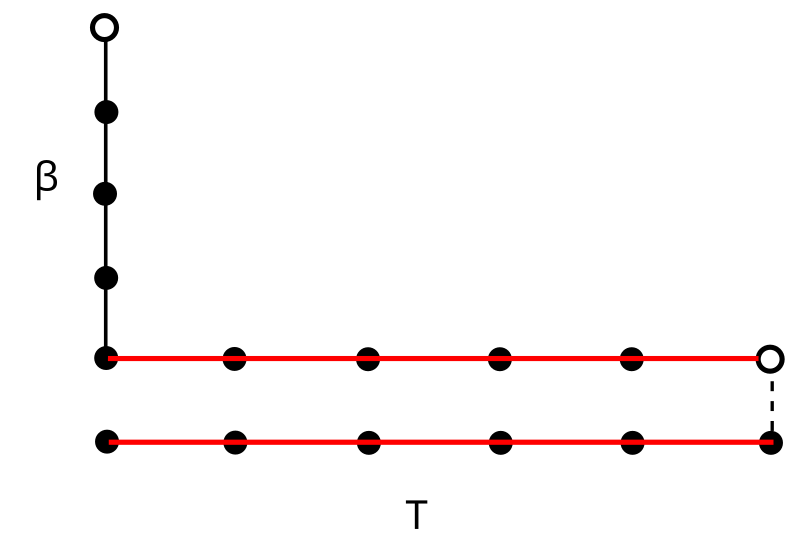}
\hfil
\includegraphics[width=0.47\linewidth]{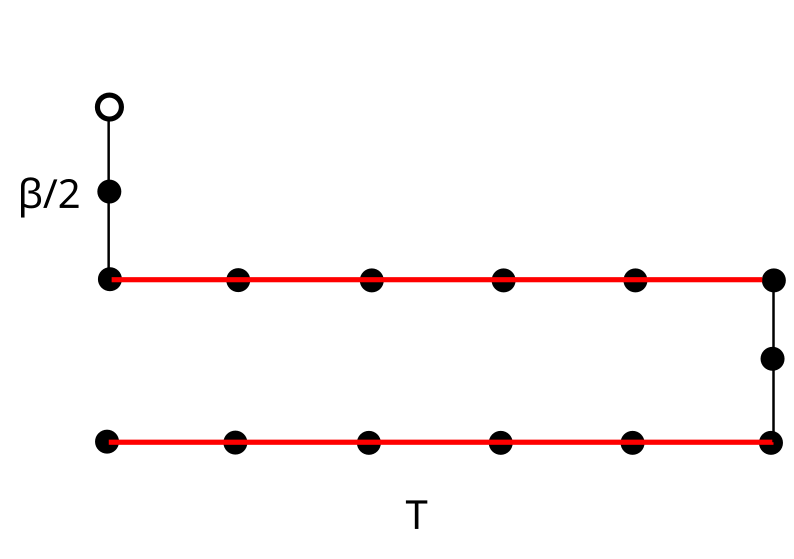}
\caption{Two common choices of Schwinger-Keldysh contour. The horizontal axis represents real-time evolution; the (periodic) vertical axis corresponds to imaginary-time evolution. The contour on the left is termed the L-contour, and the other the S-contour. Only the L-contour is useful for computing out-of-time-order correlations, or quenches.\label{fig:sk}}
\end{figure}

The choice of contour is largely arbitrary, with two major restrictions. First, if we want to compute the time-separated correlator $\langle \mathcal O(T) \mathcal O(0)\rangle$, the points $t=0$ and $t=T$ must be included on the contour. Second, because $H$ is typically unbounded above, the operator $e^{-i H t}$ is defined only on the lower half-plane; therefore the chosen contour must never move `backward' in imaginary time. As a consequence of these two constraints, every contour that allows us to compute $\langle \mathcal O(T) \mathcal O(0)\rangle$ must include the straight line from $t=0$ to $t=T$. The two most common contours are shown in Figure~\ref{fig:sk}. Except where noted, in this paper we will always work with the second, the so-called S-contour.

The anharmonic oscillator is of course just the $0+1$-dimensional case of $\phi^4$ field theory, defined by the lattice action
\begin{align}
S_{\phi^4} = &\sum_{t,x} \frac{|\phi(t,x)-\phi(t+1,x)|^2}{2a(t)} +\\ &\sum_{t,x}\frac{a(t)+a(t-1)}{2}\left[\frac{|\phi(t,x)-\phi(t,x+1)|^2}{2} + \frac{m^2}{2} |\phi(t,x)|^2+ \lambda |\phi(t,x)|^4\right]
\text,
\end{align}
where we have defined $|\phi(t,x)|^2 \equiv \sum_i \phi_i(t,x)^2$ for brevity. The spatial lattice spacing is implicitly made to be $1$.
The same considerations of the choice of contour $a(t)$ apply to scalar field theory, as well as to the fermionic theories below. Note that the spatial lattice spacing is assumed here and throughout to be $1$, while the time-like spacing may be different. Usually, the time-like spacing is smaller, to approximate the Hamiltonian limit.

A fermionic analogue to $\phi^4$ scalar field theory is given by the Gross-Neveu model, described in the continuum by the action
\begin{equation}\label{eq:S_GN}
S_{GN} = \int \! d^d x\; \bar\psi \big(\slashed\partial + m\big)\psi + \frac{g^2}{2N} (\bar \psi \psi )^2
\text.
\end{equation}
Modifying the interaction term to $(\bar\psi \gamma^\mu \psi)^2$ yields the Thirring model; for $d=2$ and $N=1$, the two models are equivalent (up to a redefinition of the coupling $g$).

\section{Large $N$ on the Lattice}

For all models discussed in the section above, expectation values in or near the large-$N$ limit can be evaluated as a saddle point expansion. This is accomplished by introducing an auxiliary field $\zeta$ to absorb the quartic interaction term. Suitably normalizing the field $\zeta$ results in the action appearing with $N$ as an overall factor. Thus, in the large-$N$ limit, the path integral is dominated by the saddle points (in practice, one particular saddle point) of the per-flavor action.

In the case of scalar field theory in $d$ spatial dimensions, the resulting effective action reads
\begin{equation}
S_{\phi^4,\mathrm{eff}}[\zeta] = \frac{N}{16 \lambda} \sum_{x,t} a(t) \zeta(t,x)^2 + \frac N 2 \log \det M(\zeta)
\text,
\end{equation}
where the inverse propagator in a background field of $\zeta$ is given, in terms of the free lattice boson inverse propagator $M^{(0)}$, by
\begin{equation}
M_{xy} = M^{(0)}_{xy} + \delta_{xy} \big(m^2 + i \zeta(x)\big)\text.
\end{equation}
This effective action is valid in any number of dimensions; in particular the anharmonic oscillator is obtained as the $d=0$ case.

In the case of the fermionic models, a lattice calculation even at finite $N$ already requires a similar procedure, in which an auxiliary field is introduced and the fermions are integrated out to yield a fermion determinant --- this time in the numerator. It can be arranged for the number of flavors $N$ to appear as a prefactor on the entire action, so that the effective action is of the form
\begin{equation}
S_{F,\mathrm{eff}}[\sigma] = \frac{N}{8 g^2} \sum a(t) \zeta(t)^2 - N \log \det D(\zeta)
\text,
\end{equation}
where $g^2$ is the lattice coupling constant, and $D(\zeta)$ is the Dirac operator in a non-dynamical background field $\zeta$. (This form is the same whether the Gross-Neveu or Thirring model is being studied; only the Dirac operator differs.)

This way of rewriting the theories exactly mirrors the procedure used to address the large-$N$ limit in analytic, continuum results. The next step is also little different: the path integral, which in both cases is of the form
\begin{equation}\label{eq:path-integral}
Z = \int \mathcal D \zeta\; e^{-N S_{\mathrm{eff},1}(\zeta)}
\text,
\end{equation}
where $S_{\mathrm{eff},1}$ is the per-flavor action, can be approximated by an expansion around the saddle point. This yields the $\frac 1 N$ expansion for all observables.

\section{Contour Deformations}

The real-time components of the Schwinger-Keldysh contour contribute a sign problem. The ``Boltzmann factor'' $e^{-S}$ is complex, and can no longer be treated as a probability distribution. Sign problems can be handled, to some extent, by reweighting: sampling with respect to the quenched distribution $|e^{-S}|$, and computing the desired expectation value as a ratio of quenched expectation values. This procedure converges exponentially slowly in the lattice volume, and sometimes not at all.

A general approach to alleviating sign problems, first proposed in~\cite{Cristoforetti:2012su}, is to treat the path integral (\ref{eq:path-integral}) as a complex contour integral along $\mathbb R^V \subset \mathbb C^V$, and deform the contour of integration to a submanifold of $\mathbb C^V$ with a milder sign problem. As long as the Boltzmann factor is a holomorphic function of the auxiliary fields, physical observables are not changed by this process. (See~\cite{Alexandru:2020wrj} for a modern review of this family of methods.)

In the bosonic case, a difficulty immediately arises: the integrand of the path integral is not, in fact, holomorphic. The propagator, which appears in the denominator, can be singular, introducing poles. For this reason, we are prevented from performing a proper Monte Carlo over configurations for bosonic field theory. Nevertheless, the saddle point expansion still yields realistic and useful results at leading order and $N^{-1}$.

The propagator in the fermionic case can still be singular, but this introduces only zeros into the integrand of the path integral, and no poles. As a result, deformations of the integration contour are guaranteed not to change any expectation values\footnote{Note that correlators of the fermion fields involve inverses of the Dirac operator, but as shown in~\cite{Alexandru:2018ngw} the resulting singularities are exactly cancelled by the factor of $\det D$ in the Boltzmann factor, protecting the holomorphicity of the integrand.}, and a Monte Carlo on the tangent plane is trustworthy.

\section{Lattice Calculation}

\begin{figure}
\centering
\includegraphics[width=0.46\textwidth]{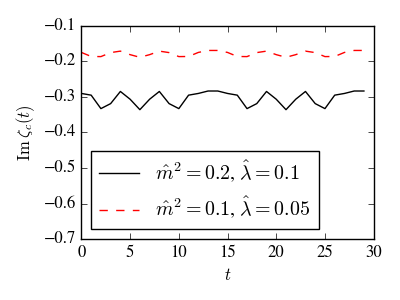}
\caption{The saddle point for the anharmonic oscillator (\ref{eq:H_aho}). The imaginary part of the saddle point $\zeta_c(t)$ is plotted as a function of the lattice time-slice, for two different sets of bare parameters; the real part vanishes identically. Closer to the continuum limit, the fluctuations in $\zeta_c(t)$ disappear. \label{fig:aho-saddle}}
\end{figure}

The first step required in any saddle-point based calculation is to find the saddle point. In the continuum, this is a constant field $\zeta = \zeta_0$; however, lattice artifacts introduce small fluctuations. To make a numerical search practical, we use the fact that the saddle point is pure-imaginary, and that for the S-contour, it obeys the symmetry $\zeta(t) = \zeta(\hat \beta + 2\hat T - t)^\dagger$.  Figure~\ref{fig:aho-saddle} shows the saddle point $i\zeta(t)$ of the anharmonic oscillator for $\hat \beta = 6$, $\hat T = 12$, for two different values of the bare parameters $m$ and $\lambda$.

The most straightforward way to proceed with a lattice calculation is to perform a Monte Carlo, at fixed but large $N$, on the tangent plane to the dominant thimble. At large $N$, the probability distribution on the tangent place is approximately Gaussian, and can be sampled efficiently (with moderate reweighting) without the need of a Markov Chain. At slightly smaller $N$, a Markov Chain becomes helpful, although reweighting is still required to address the mild sign problem that develops at any finite $N$.

As discussed in the previous section, direct sampling of field configurations fails in the presence of singularities of the Boltzmann factor, which occur for the scalar theories. Alternatively, terms of the expansion in $N^{-1}$ can be computed explicitly. As an example, consider the two-point function $\langle \phi(t) \phi(0)\rangle$. The leading-order piece is just that expectation value evaluated in the free field in a background of $\zeta_c$. The next contribution is at order $N^{-1}$, and is naturally expressed in terms correlators of a rescaled field $\sigma = \sqrt N \left(\zeta-\zeta_c\right)$.
\begin{align}
\frac{d}{d N^{-1}} \langle \phi(t) \phi(0) \rangle
&=
\big\langle
\Big(\frac 1 6 A_{xyz} \Sigma_{xyz}\Big)^2 M^{-1}_{t0}
- \frac 1 {24} B_{wxyz} \Sigma_{wxyz} M^{-1}_{t0}
\big\rangle_0
\nonumber\\
&+
\big\langle
\frac i 6 A_{xyz} \Sigma_{xyz} a_w M^{-1}_{xw} M^{-1}_{w0}
- a_y a_z M^{-1}_{xy} M^{-1}_{yz} M^{-1}_{z0}
\big\rangle_0
\nonumber\\
&-
\big\langle M^{-1}_{t0} \big\rangle_0
\big\langle \Big(\frac 1 6 A_{xyz} \Sigma_{xyz}\Big)^2 - \frac 1 {24} B_{wxyz} \Sigma_{wxyz}\big\rangle_0
\text,\label{eq:nlo}
\end{align}
where $\langle \cdot\rangle_0$ denotes a leading-order expectation value --- a Gaussian expectation value evaluated at the saddle point. The tensors $M$, $A$, and $B$ give the second, third, and fourth derivatives of the effective action with respect to the fields $\sigma$:
\begin{equation}
S_{\mathrm{eff}}(\sigma) = M_{xy} \Sigma_{xy} + \frac 1 {\sqrt N} A_{xyz} \Sigma_{xyz} + \frac 1 N B_{wxyz} \Sigma_{wxyz}\text.
\end{equation}
For brevity, here and above, we have used the notation $\Sigma_{xy} = \sigma_x \sigma_y$ (and similarly with three and four indices). Finally, $a_x$ gives the direction of the Schwinger-Keldysh contour at site $x$. Where an index appears thrice in (\ref{eq:nlo}), it is summed over just as if it had appeared only twice.

The required Gaussian expectation values can be obtained semi-analytically, by summing over contractions, or via a Monte Carlo, by sampling from the Gaussian distribution about the saddle point. Either method requires exponential time in the order of the expansion desired.

\begin{figure}
\hfil
\includegraphics[width=0.46\textwidth]{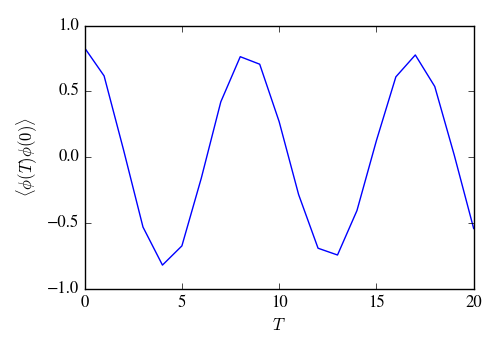}
\hfil
\includegraphics[width=0.46\textwidth]{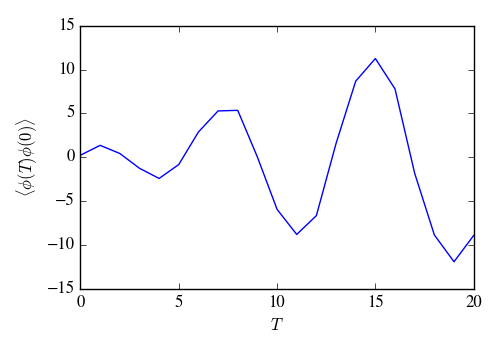}
\hfil
\caption{Large-$N$ expansion of the $2$-point function of the anharmonic oscillator, with $m^2 = 0.2$, $\lambda = 0.1$. The panel on the left shows the leading-order term; i.e., $\langle \phi(t)\phi(0)\rangle$ in the large-$N$ limit. On the right is the $N^{-1}$ contribution.\label{fig:scalar}}
\end{figure}

Figure~\ref{fig:scalar} shows the two-point function $\langle \phi_i(t)\phi_i(0)\rangle$ in the large-$N$ expansion. This correlator is a simple sinusoid, with frequency giving the mass of the lowest lying spacetime scalar, $O(N)$ vector particle. Writing the mass of this particle as $m(N) = m_0 + m_2 N^{-1} + O(N^{-2})$, the correlator has the form
\begin{equation}
\langle \phi(t) \phi(0) \rangle \sim \cos \big(m_0 + m_2 N^{-1}\big)t
\approx \cos m_0 t - N^{-1} m_2 t \sin m_0 t
\text.
\end{equation}
Thus, the large-$N$ limit $m_0$ of the mass is given by the frequency of the left panel, and the subleading term $m_2$ is given by the slope of the amplitude of the sinusoid in the right panel of figure~\ref{fig:scalar}.

The $\frac 1 N$ corrections to the two-point function, of course, are readily obtained analytically. A key advantage of the lattice formulation is enabling the calculation of observables which are not so analytically tractable. One example is the \emph{quench} --- a system is permitted to equilibrate under one Hamiltonian $H_0$, and then at some time $t=0$, the Hamiltonian is changed to $H_1$.

A nontrivial quench can be achieved by a change of the coupling $\lambda$. Note that a quench is whicht he coupling is turned \emph{off} --- set to $\lambda=0$ for the real-time evolution --- is both less interesting and easier to computer. The real-time part of the Schwinger-Keldysh contour for such a quench corresponds to evolution of the free theory, which is efficiently simulable on a classical computer.

\section{Discussion}

The large-$N$ limit of $O(N)$ symmetric bosons and $U(N)$ symmetric fermions is amenable to efficient calculations either analytically or on the lattice. The lattice expansion makes accessible more complex physics, including quenches, that are analytically tractable. In cases where the lattice would ordinarily have a sign problem, the large-$N$ limit renders Lefschetz thimble-based methods perfectly effective in resolving that sign problem.

These calculations are non-perturbative in the sense that all orders of the bare coupling $\lambda$ are involved in the particle masses, equation of state, and so on. However, the S-matrix itself is very close to the identity, with leading-order deviations that are $O(N^{-1})$. From a standpoint of computational complexity, this is the property that enables the calculations performed above. In the $\frac 1 N$ expansion, particles cannot interact an arbitrary number of times, preventing the simulation of arbitrary quantum circuits at any fixed order in $N^{-1}$. The size of circuit that can be simulated is proportional to the order of the expansion; correspondingly, the calculations above are exponentially difficult in the order of the expansion. In other words, these calculations are hard exactly where they `should' be to avoid an efficient classical algorithm for simulating arbitrary quantum dynamics.

\acknowledgments
I am grateful to Paul Romatschke and Max Weiner for many useful conversations. This material is based upon work supported by the U.S. Department of Energy, Office of Science, Office of Nuclear Physics program under Award Number DE-SC-0017905.

\bibliographystyle{JHEP}
\bibliography{refs}

\end{document}